\documentclass[twocolumn,letterpaper,aps,prc,nofootinbib,longbibliography,superscriptaddress,showpacs,floatfix]{revtex4-1}

\usepackage{graphicx}	
\usepackage{amsmath,amssymb,amsfonts}
\usepackage{xspace}	

\newcommand{\pt}{\mbox{$p_T$}\xspace}

\newcommand{\KEt}{\mbox{${\rm KE}_T$}\xspace}

\def\fig#1{{Fig.~\ref{#1}}}

\newcommand{\mean}[1]{\left\langle #1 \right\rangle}

\begin{document}

\title{Measurement of the higher-order anisotropic flow coefficients 
       for identified hadrons in Au$+$Au collisions 
       at $\sqrt{s_{_{NN}}}$ = 200 GeV}

\newcommand{\abilene}{Abilene Christian University, Abilene, Texas 79699, USA}
\newcommand{\augie}{Department of Physics, Augustana University, Sioux Falls, South Dakota 57197, USA}
\newcommand{\banaras}{Department of Physics, Banaras Hindu University, Varanasi 221005, India}
\newcommand{\barc}{Bhabha Atomic Research Centre, Bombay 400 085, India}
\newcommand{\baruch}{Baruch College, City University of New York, New York, New York, 10010 USA}
\newcommand{\bnlcoll}{Collider-Accelerator Department, Brookhaven National Laboratory, Upton, New York 11973-5000, USA}
\newcommand{\bnlphys}{Physics Department, Brookhaven National Laboratory, Upton, New York 11973-5000, USA}
\newcommand{\caucr}{University of California - Riverside, Riverside, California 92521, USA}
\newcommand{\charlesczech}{Charles University, Ovocn\'{y} trh 5, Praha 1, 116 36, Prague, Czech Republic}
\newcommand{\chonbuk}{Chonbuk National University, Jeonju, 561-756, Korea}
\newcommand{\ciae}{Science and Technology on Nuclear Data Laboratory, China Institute of Atomic Energy, Beijing 102413, People's~Republic~of~China}
\newcommand{\cns}{Center for Nuclear Study, Graduate School of Science, University of Tokyo, 7-3-1 Hongo, Bunkyo, Tokyo 113-0033, Japan}
\newcommand{\colorado}{University of Colorado, Boulder, Colorado 80309, USA}
\newcommand{\columbia}{Columbia University, New York, New York 10027 and Nevis Laboratories, Irvington, New York 10533, USA}
\newcommand{\czechtech}{Czech Technical University, Zikova 4, 166 36 Prague 6, Czech Republic}
\newcommand{\dapnia}{Dapnia, CEA Saclay, F-91191, Gif-sur-Yvette, France}
\newcommand{\debrecen}{Debrecen University, H-4010 Debrecen, Egyetem t{\'e}r 1, Hungary}
\newcommand{\elte}{ELTE, E{\"o}tv{\"o}s Lor{\'a}nd University, H-1117 Budapest, P\'azmany P\'eter s\'et\'any 1/A, Hungary}
\newcommand{\ewha}{Ewha Womans University, Seoul 120-750, Korea}
\newcommand{\fit}{Florida Institute of Technology, Melbourne, Florida 32901, USA}
\newcommand{\fsu}{Florida State University, Tallahassee, Florida 32306, USA}
\newcommand{\gsu}{Georgia State University, Atlanta, Georgia 30303, USA}
\newcommand{\hiroshima}{Hiroshima University, Kagamiyama, Higashi-Hiroshima 739-8526, Japan}
\newcommand{\ihepprot}{IHEP Protvino, State Research Center of Russian Federation, Institute for High Energy Physics, Protvino, 142281, Russia}
\newcommand{\illuiuc}{University of Illinois at Urbana-Champaign, Urbana, Illinois 61801, USA}
\newcommand{\inrras}{Institute for Nuclear Research of the Russian Academy of Sciences, prospekt 60-letiya Oktyabrya 7a, Moscow 117312, Russia}
\newcommand{\instpasczech}{Institute of Physics, Academy of Sciences of the Czech Republic, Na Slovance 2, 182 21 Prague 8, Czech Republic}
\newcommand{\isu}{Iowa State University, Ames, Iowa 50011, USA}
\newcommand{\jaea}{Advanced Science Research Center, Japan Atomic Energy Agency, 2-4 Shirakata Shirane, Tokai-mura, Naka-gun, Ibaraki-ken 319-1195, Japan}
\newcommand{\jinrdubna}{Joint Institute for Nuclear Research, 141980 Dubna, Moscow Region, Russia}
\newcommand{\jyvaskyla}{Helsinki Institute of Physics and University of Jyv{\"a}skyl{\"a}, P.O.Box 35, FI-40014 Jyv{\"a}skyl{\"a}, Finland}
\newcommand{\kek}{KEK, High Energy Accelerator Research Organization, Tsukuba, Ibaraki 305-0801, Japan}
\newcommand{\korea}{Korea University, Seoul, 136-701, Korea}
\newcommand{\kurchatov}{National Research Center ``Kurchatov Institute," Moscow, 123098 Russia}
\newcommand{\kyoto}{Kyoto University, Kyoto 606-8502, Japan}
\newcommand{\labllr}{Laboratoire Leprince-Ringuet, Ecole Polytechnique, CNRS-IN2P3, Route de Saclay, F-91128, Palaiseau, France}
\newcommand{\lahorelums}{Physics Department, Lahore University of Management Sciences, Lahore 54792, Pakistan}
\newcommand{\lawllnl}{Lawrence Livermore National Laboratory, Livermore, California 94550, USA}
\newcommand{\losalamos}{Los Alamos National Laboratory, Los Alamos, New Mexico 87545, USA}
\newcommand{\lpc}{LPC, Universit{\'e} Blaise Pascal, CNRS-IN2P3, Clermont-Fd, 63177 Aubiere Cedex, France}
\newcommand{\lund}{Department of Physics, Lund University, Box 118, SE-221 00 Lund, Sweden}
\newcommand{\maryland}{University of Maryland, College Park, Maryland 20742, USA}
\newcommand{\mass}{Department of Physics, University of Massachusetts, Amherst, Massachusetts 01003-9337, USA }
\newcommand{\michigan}{Department of Physics, University of Michigan, Ann Arbor, Michigan 48109-1040, USA}
\newcommand{\muenster}{Institut f\"ur Kernphysik, University of Muenster, D-48149 Muenster, Germany}
\newcommand{\muhlenberg}{Muhlenberg College, Allentown, Pennsylvania 18104-5586, USA}
\newcommand{\myongji}{Myongji University, Yongin, Kyonggido 449-728, Korea}
\newcommand{\nagasaki}{Nagasaki Institute of Applied Science, Nagasaki-shi, Nagasaki 851-0193, Japan}
\newcommand{\nara}{Nara Women's University, Kita-uoya Nishi-machi Nara 630--8506, Japan}
\newcommand{\natmephi}{National Research Nuclear University, MEPhI, Moscow Engineering Physics Institute, Moscow, 115409, Russia}
\newcommand{\newmex}{University of New Mexico, Albuquerque, New Mexico 87131, USA }
\newcommand{\nmsu}{New Mexico State University, Las Cruces, New Mexico 88003, USA}
\newcommand{\ohio}{Department of Physics and Astronomy, Ohio University, Athens, Ohio 45701, USA}
\newcommand{\ornl}{Oak Ridge National Laboratory, Oak Ridge, Tennessee 37831, USA}
\newcommand{\orsay}{IPN-Orsay, Univ. Paris-Sud, CNRS/IN2P3, Universit{\'e} Paris-Saclay, BP1, F-91406, Orsay, France}
\newcommand{\peking}{Peking University, Beijing 100871, People's~Republic~of~China}
\newcommand{\pnpi}{PNPI, Petersburg Nuclear Physics Institute, Gatchina, Leningrad Region, 188300, Russia}
\newcommand{\riken}{RIKEN Nishina Center for Accelerator-Based Science, Wako, Saitama 351-0198, Japan}
\newcommand{\rikjrbrc}{RIKEN BNL Research Center, Brookhaven National Laboratory, Upton, New York 11973-5000, USA}
\newcommand{\rikkyo}{Physics Department, Rikkyo University, 3-34-1 Nishi-Ikebukuro, Toshima, Tokyo 171-8501, Japan}
\newcommand{\saispbstu}{Saint Petersburg State Polytechnic University, St. Petersburg, 195251 Russia}
\newcommand{\saopaulo}{Universidade de S{\~a}o Paulo, Instituto de F\'{\i}sica, Caixa Postal 66318, S{\~a}o Paulo CEP05315-970, Brazil}
\newcommand{\seoulnat}{Department of Physics and Astronomy, Seoul National University, Seoul 151-742, Korea}
\newcommand{\stonybrkc}{Chemistry Department, Stony Brook University, SUNY, Stony Brook, New York 11794-3400, USA}
\newcommand{\stonycrkp}{Department of Physics and Astronomy, Stony Brook University, SUNY, Stony Brook, New York 11794-3800, USA}
\newcommand{\tenn}{University of Tennessee, Knoxville, Tennessee 37996, USA}
\newcommand{\titech}{Department of Physics, Tokyo Institute of Technology, Oh-okayama, Meguro, Tokyo 152-8551, Japan}
\newcommand{\tsukuba}{Institute of Physics, University of Tsukuba, Tsukuba, Ibaraki 305, Japan}
\newcommand{\vandy}{Vanderbilt University, Nashville, Tennessee 37235, USA}
\newcommand{\waseda}{Waseda University, Advanced Research Institute for Science and Engineering, 17 Kikui-cho, Shinjuku-ku, Tokyo 162-0044, Japan}
\newcommand{\weizmann}{Weizmann Institute, Rehovot 76100, Israel}
\newcommand{\wigner}{Institute for Particle and Nuclear Physics, Wigner Research Centre for Physics, Hungarian Academy of Sciences (Wigner RCP, RMKI) H-1525 Budapest 114, POBox 49, Budapest, Hungary}
\newcommand{\yonsei}{Yonsei University, IPAP, Seoul 120-749, Korea}
\affiliation{\abilene}
\affiliation{\augie}
\affiliation{\banaras}
\affiliation{\barc}
\affiliation{\baruch}
\affiliation{\bnlcoll}
\affiliation{\bnlphys}
\affiliation{\caucr}
\affiliation{\charlesczech}
\affiliation{\chonbuk}
\affiliation{\ciae}
\affiliation{\cns}
\affiliation{\colorado}
\affiliation{\columbia}
\affiliation{\czechtech}
\affiliation{\dapnia}
\affiliation{\debrecen}
\affiliation{\elte}
\affiliation{\ewha}
\affiliation{\fit}
\affiliation{\fsu}
\affiliation{\gsu}
\affiliation{\hiroshima}
\affiliation{\ihepprot}
\affiliation{\illuiuc}
\affiliation{\inrras}
\affiliation{\instpasczech}
\affiliation{\isu}
\affiliation{\jaea}
\affiliation{\jinrdubna}
\affiliation{\jyvaskyla}
\affiliation{\kek}
\affiliation{\korea}
\affiliation{\kurchatov}
\affiliation{\kyoto}
\affiliation{\labllr}
\affiliation{\lahorelums}
\affiliation{\lawllnl}
\affiliation{\losalamos}
\affiliation{\lpc}
\affiliation{\lund}
\affiliation{\maryland}
\affiliation{\mass}
\affiliation{\michigan}
\affiliation{\muenster}
\affiliation{\muhlenberg}
\affiliation{\myongji}
\affiliation{\nagasaki}
\affiliation{\nara}
\affiliation{\natmephi}
\affiliation{\newmex}
\affiliation{\nmsu}
\affiliation{\ohio}
\affiliation{\ornl}
\affiliation{\orsay}
\affiliation{\peking}
\affiliation{\pnpi}
\affiliation{\riken}
\affiliation{\rikjrbrc}
\affiliation{\rikkyo}
\affiliation{\saispbstu}
\affiliation{\saopaulo}
\affiliation{\seoulnat}
\affiliation{\stonybrkc}
\affiliation{\stonycrkp}
\affiliation{\tenn}
\affiliation{\titech}
\affiliation{\tsukuba}
\affiliation{\vandy}
\affiliation{\waseda}
\affiliation{\weizmann}
\affiliation{\wigner}
\affiliation{\yonsei}
\author{A.~Adare} \affiliation{\colorado}
\author{S.~Afanasiev} \affiliation{\jinrdubna}
\author{C.~Aidala} \affiliation{\mass} \affiliation{\michigan}
\author{N.N.~Ajitanand} \affiliation{\stonybrkc}
\author{Y.~Akiba} \affiliation{\riken} \affiliation{\rikjrbrc}
\author{H.~Al-Bataineh} \affiliation{\nmsu}
\author{J.~Alexander} \affiliation{\stonybrkc}
\author{K.~Aoki} \affiliation{\kyoto} \affiliation{\riken}
\author{Y.~Aramaki} \affiliation{\cns}
\author{E.T.~Atomssa} \affiliation{\labllr}
% ...
\collaboration{PHENIX Collaboration} \noaffiliation

\date{\today}



\begin{abstract}


Measurements of the anisotropic flow coefficients 
$v_2\{\Psi_2\}$, $v_3\{\Psi_3\}$, $v_4\{\Psi_4\}$, and 
$v_4\{\Psi_2\}$ for identified particles ($\pi^{\pm}$, $K^{\pm}$, 
and $p+\bar{p}$) at midrapidity, obtained relative to the event 
planes $\Psi_m$ at forward rapidities in Au$+$Au collisions at 
$\sqrt{s_{_{NN}}}$~=~200 GeV, are presented as a function of 
collision centrality and particle transverse momenta $p_T$. The 
$v_n$ coefficients show characteristic patterns consistent with 
hydrodynamical expansion of the matter produced in the 
collisions.  For each harmonic $n$, a modified valence quark 
number $N_q$ scaling (plotting $v_n\{\Psi_m\}/(N_q)^{n/2}$ versus 
${\rm KE}_T/N_q$) is observed to yield a single curve for all the 
measured particle species for a broad range of transverse kinetic 
energies ${\rm KE}_T$. A simultaneous blast-wave model fit to the 
observed $v_n\{\Psi_m\}(p_T)$ coefficients and published particle spectra 
identifies radial flow anisotropies $\rho_n\{\Psi_m\}$ and spatial 
eccentricities $s_n\{\Psi_m\}$ at freeze-out.  These are generally smaller 
than the initial-state participant-plane (PP) geometric eccentricities 
$\varepsilon_n\{\Psi_m^{\rm PP}\}$, 
as also observed in the final eccentricity from quantum 
interferometry measurements with respect to the event plane.

\end{abstract}

	
\maketitle



{\it Introduction}.
The quark-gluon plasma (QGP) is a novel phase of nuclear matter 
at high temperature and energy density, whose existence is 
predicted by quantum chromodynamics~\cite{Shuryak:1983zb}. A wide 
variety of experimental observations at the Relativistic Heavy 
Ion Collider 
(RHIC)~\cite{Arsene:2004fa,Back:2004je,Adams:2005dq,Adcox:2004mh} 
provide strong evidence for the formation of a QGP in 
ultra-relativistic heavy ion collisions, particularly (1)~the 
magnitude of the observed suppression of high-\pt 
($p_T\agt4$~GeV/$c$) particles, relative to the scaled yield from 
$p$$+$$p$ collisions; and (2)~the large azimuthal anisotropy 
or anisotropic flow of the low-\pt ($p_T\alt$~3--4~GeV/$c$) bulk 
of hadrons in the final state.  The flow of low-\pt particles has 
been attributed to anisotropic expansion of the 
QGP~\cite{Song:2010mg,Denicol:2010tr,Schenke:2011bn}, and 
consequently the measured strength of anisotropic flow should be 
sensitive to the transport properties of the QGP and the 
mechanism for its space-time evolution.

The magnitude of anisotropic flow can be quantified by the 
Fourier coefficients 
$v_n\{\Psi_m\}=\mean{\cos(n(\phi-\Psi_{m}))}$ of the azimuthal 
distribution of produced 
particles~\cite{Voloshin:1994mz,Poskanzer:1998yz,Ollitrault:1992bk,Adare:2010ux}, 
where $n$ and $m$ are the order of the harmonics, $\phi$ is the 
azimuthal angle of the particles, and $\Psi_m$ is the azimuthal 
angle of the $m^{th}$ order event plane. In early studies with 
symmetric systems, $v_n\{\Psi_m\}$ was presumed to be zero for 
odd $n$ owing to the assumption that initial-state energy 
densities were smooth and symmetric across the transverse plane. 
The recent observations of sizable $v_n\{\Psi_n\}$ values for odd 
$n$~\cite{Adare:2011tg,ALICE:2011ab,ATLAS:2012at,Chatrchyan:2012wg,Adamczyk:2013waa} 
confirms the important role of fluctuations in the initial-state 
collision geometry~\cite{Alver:2010gr}.

Model-dependent analyses of higher-order harmonics for inclusive 
hadrons measured in Au$+$Au and Pb$+$Pb collisions at RHIC and 
the Large Hadron Collider have indicated that such measurements 
can provide simultaneous constraints for initial-state 
fluctuation models and the ratio of shear viscosity to 
entropy density of the 
QGP~\cite{Adare:2011tg,Schenke:2011bn,Lacey:2011ug,Gardim:2012yp}. 
The new data on higher-order $v_n\{\Psi_m\}$ for identified particles 
presented here provides additional information about the 
initial conditions and hydrodynamic properties. Here, we show 
that our $v_n\{\Psi_m\}$ measurements for different particle species 
provide (1) further tests for the constituent quark number 
scaling and quark coalescence 
models~\cite{Greco:2003xt,Fries:2003vb,Molnar:2003ff} by 
extending our previously observed scaling for 
$v_2\{\Psi_2\}$~\cite{Adare:2006ti,Adare:2014bga} to higher harmonics~\cite{Han:2011iy}; and 
(2) freeze-out parameters for hydrodynamic expansion with 
anisotropic blast-wave (BW) model 
fits~\cite{Schnedermann:1993ws,Huovinen:2001cy,Adler:2001nb,Masui:2007sca}.


{\it Data taking and particle identification}.
The results presented here for Au$+$Au collisions at 
$\sqrt{s_{_{NN}}}~=~200$~GeV are obtained with the PHENIX 
experiment from an analysis of $4.14 \times 10^9$ minimum-bias 
events taken during the 2007 running period.  Collision centrality 
is determined with the beam-beam counters~\cite{phenixDect}. 
Charged hadrons are reconstructed in a pseudorapidity ($\eta$) 
range of $\left|\eta\right|<0.35$ using the drift-chamber and 
pad-chamber subsystems~\cite{Adcox:2003zp}, which achieve the 
momentum resolution $\delta p/p\approx1.3\% \oplus 1.2\% \times 
p$~(GeV/$c$)~\cite{Adare:2012vq}. The ring imaging \v{C}erenkov 
counter is employed to veto conversion electrons. 
Time-of-flight detectors in both the east (TOFE, 
$\Delta\varphi=\pi/4$ rad) and west (TOFW, $\Delta\varphi=0.342$ 
rad) arms are used for $\pi^{\pm}, K^{\pm}$, and $p+\bar{p}$ 
identification after the conversion electron 
veto~\cite{Adare:2012vq}. The timing resolution of TOFE (TOFW) is 
133 ($84\pm1$) ps. For \mbox{$\pt<3$~GeV/$c$} both TOFE and TOFW 
detectors were used. 
For \mbox{$\pt>3$~GeV/$c$} particle identification utilizes the
TOFW in conjunction with the Aerogel \v{C}erenkov Counter (ACC).  
The two detectors have a common azimuthal acceptance of 
$\Delta\varphi=0.171$~rad.
With these detectors, a $p+\bar{p}$ purity of 
greater than 97\% was achieved for \mbox{$\pt<4$~GeV/$c$}; and 
purity for $\pi^{\pm}$ and $K^{\pm}$ greater than 98\% for 
\mbox{$\pt<3$~GeV/$c$} and 90\% for \mbox{$3<\pt<4$~GeV/$c$} were 
also achieved, as detailed in~\cite{Adare:2012vq}. The 
purity and efficiency of particle identification (PID) 
are independent of the relative 
azimuthal angle between particles and the event plane 
$\phi-\Psi_m$.

	
{\it Experimental technique}.
Measurements of the flow coefficients $v_2\{\Psi_2\}$, 
$v_3\{\Psi_3\}$, $v_4\{\Psi_4\}$, and $v_4\{\Psi_2\}$ as a 
function of centrality and \pt for $\pi^{\pm}$, $K^{\pm}$, and 
$p+\bar{p}$ ({\em i.e.} with charge signs combined) are obtained 
with both the event plane (EP) and the long-range two-particle 
correlation (2PC) methods. In the EP method, a measured event 
plane direction $\Psi^{\rm obs}_m$ is determined for every event 
and for each order $m$, using the south and north reaction-plane 
detectors (RXN), covering $\Delta\varphi=2\pi$ and 
$1<|\eta|<2.8$~\cite{Richardson:2010hm}. Each is made of plastic 
scintillator paddles with lead converter in front and with 
optical fibers guided to photo multiplier tubes.  Each RXN detector is  
segmented into 12 sections in $\varphi$ and two rings in $\eta$. 
The $\Psi^{\rm obs}_{m}$ are determined via a sum over the 
azimuthal angle $\phi_i$ of each RXN element in both the arms 
with its charge $w_i$ deposited by particles for that event, as 
\mbox{$\tan(m\Psi^{\rm obs}_m) 
=\sum_{i}w_i\sin{(m\phi_i)}/\sum_{i}w_i\cos{(m\phi_i)}$}. The 
flow magnitudes \mbox{$v_n\{\Psi_m\}=\mean{\cos{n 
(\phi-\Psi^{\rm obs}_m)}}/{\rm Res}\{n,\Psi_m\}$} are then 
measured with respect to each harmonic event plane, where $\phi$ 
is the azimuthal angle of the hadron and 
${\rm Res}\{n,\Psi_m\}=\mean{\cos{n (\Psi_m-\Psi^{\rm obs}_m)}}$ 
is the event plane resolution, which is estimated for each 
centrality by the standard sub-event method as described 
in~\cite{Afanasiev:2009wq,Poskanzer:1998yz,Tsukuba:20132014dt}. 
The best resolution of each harmonic is measured to be
${\rm Res}\{2,\Psi_2\}\sim0.75$ and ${\rm 
Res}\{4,\Psi_2\}\sim0.5$ (${\rm Res}\{3,\Psi_3\}\sim0.3$ and 
${\rm Res}\{4,\Psi_4\}\sim0.15$) in 20\%--30\% (0\%--10\%) 
central collisions.

The 2PC method pairs the hadrons (HAD) with 
deposited charges in the RXN segments. 
The distribution of the relative azimuthal angles of 
particle hits in separate $\eta$ ranges $A$ and $B$, 
$\Delta\phi\equiv\phi^{A}-\phi^{B}$, reflects the product of the 
$v_n$'s via
$dN/d\Delta\phi \propto 1+\sum_{n=1}2v_n^{A}v_n^{B}\cos(n\Delta\phi)$~\cite{Poskanzer:1998yz,Lacey:2001va,Adcox:2002ms}.
We analyze the $\Delta\phi$ correlations using the mixed-event technique 
for two pair combinations; $(A,B)$=(HAD,RXN) and 
$(A,B)$=(RXN-N,RXN-S). These correlations then fix the 
event-averaged products
$\mean{v^{\rm HAD}_n v^{\rm RXN}_n}$ and 
$\mean{v^{\rm RXN}_n v^{\rm RXN}_n}$, and 
allow us to obtain 
$v_n^{HAD}=\mean{v_n^{HAD}v_n^{\rm RXN}}/\sqrt{\mean{v_n^{\rm RXN}v_n^{\rm RXN}}}$. 
Note that flow harmonics extracted with the 2PC method are not measured 
with respect to event planes.  Thus, from this point forward we refer 
to flow harmonics in the 2PC methods as $v_n\{{\rm 2PC}\}$.  We 
use $v_n$ in cases when the discussion is generically about 
either method. In both of the analysis methods used, the results 
for wider centrality ranges are obtained by averaging across
several smaller ranges, weighted
by the multiplicity of the selected particle~\cite{Masui:2012zh}.

The systematic uncertainties in the $v_n$ measurements were estimated for: 
(1)~$\eta$ acceptance variation of the RXNs, in the EP and 2PC methods; 
this is correlated among $v_n(\pt)$ for each hadron species with the same 
fractional $v_n$ amount in the entire \pt range, except for 
$v_4\{\Psi_4\}$ where it tends to decrease as \pt increases;
(2)~detector acceptance effects of TOFE and TOFW, including occupancy; these are correlated 
among $v_n(\pt)$ for each hadron species with the same $v_n$ constant in the entire \pt range;
(3)~hadron track/hit matching cut; and 
(4)~particle identification purity.
The systematic uncertainties (1) and (2) are \pt-correlated, 
while (3) and (4) are \pt-uncorrelated.
These uncertainties are similar between the EP and 2PC methods.
Table~\ref{tab:1} summarizes typical systematic uncertainties 
on the different $v_n\{\Psi_m\}$ measures in the EP method for $\pi^\pm$ at $\pt=2$~GeV/$c$.

 \begin{table}[tbh]
 \caption{\label{tab:1}
Systematic uncertainties on the measured $v_n\{\Psi_m\}$ by EP method for $\pi^\pm$ at 
$\pt=2$~GeV/$c$ in 0\%--10\% (30\%--50\%) central collisions. 
Uncertainties of type (2) are absolute in $v_n\{\Psi_m\}$ value with the 
multiplication factor $10^{-3}$; the others are relative 
fractions of $v_n\{\Psi_m\}$ expressed in percent.
}
 \begin{ruledtabular} \begin{tabular}{cclccccc}
 & Type & Source & $v_2\left\{\Psi_2\right\}$ & $v_3\left\{\Psi_3\right\}$ & $v_4\left\{\Psi_4\right\}$ & $v_4\left\{\Psi_2\right\}$ & \\ \hline
 & (1) & RXN $\eta$[\%]         & 4.3(3.0) & 4.7(12.5) & 16(31)   &  34(7.0)  & \\
 & (2) & Acceptance[$10^{-3}$]  & 5.0(1.0) & 0.5(2.0)  & 0.7(2.5) &  0.1(0.2) & \\
 & (3) & Matching[\%]           & 1.4(0.3) & 0.7(1.0)  & 2.6(2.8) &  7.7(1.7) & \\
 & (4) & PID[\%]        & 0.3(0.1) & 0.3(0.3)  & 0.8(1.0) &  2.7(0.4) & \\
 \end{tabular} \end{ruledtabular}
 \end{table}

\begin{figure}[tbp]
\includegraphics[width=1.0\linewidth]{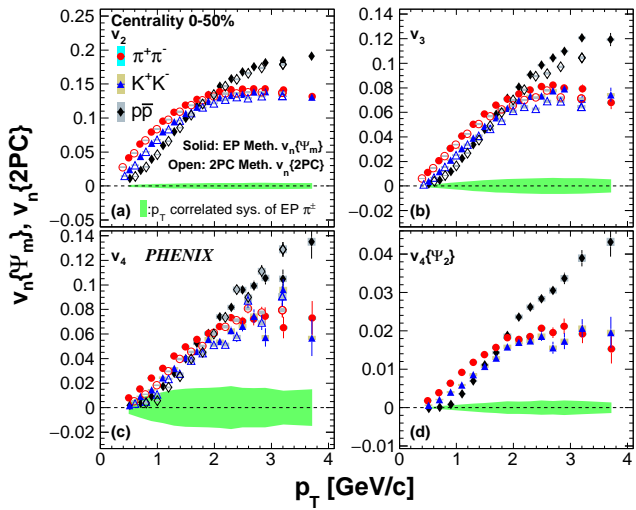}
\caption{(Color online) 
Fourier coefficients for charge-combined $\pi^{\pm}$, $K^{\pm}$, and 
$p+\bar{p}$ at midrapidity for 0\%--50\% central Au$+$Au collisions at 
$\sqrt{s_{_{NN}}}$ = 200 GeV. 
Different \pt bins were used for the EP and 2PC methods.
The green bands indicate the 
\pt-correlated systematic uncertainties of the $\pi^{\pm}$ results from 
the EP method.
The shaded boxes around the data points are 
\pt-uncorrelated systematic uncertainties, which are smaller than 
the symbols in many cases.
}
\label{fig:1}
\end{figure}


{\it Results for 0\%--50\% centrality bin}.
Figures~\ref{fig:1}(a)--(c) show a comparison of $v_2(p_T)$, 
$v_3(p_T)$, and $v_4(p_T)$ for $\pi^{\pm}$, $K^{\pm}$, and 
$p+\bar{p}$ for the EP (solid points) and 2PC (open points) 
methods in a \mbox{0\%--50\%} centrality sample; they indicate 
very good agreement between the two methods. Shown in 
Fig.~\ref{fig:1}(d) is $v_4\{\Psi_2\}$, {\em i.e.}, the fourth 
harmonic coefficient with respect to the second-order harmonic 
event plane. It can be seen that $v_4\{\Psi_2\}$ is smaller than 
$v_4\{\Psi_4\}$ but still sizable, indicating significant 
correlations between $\Psi_2$ and $\Psi_4$~\cite{Aad:2014fla}, 
which can be ascertained through the trigonometric identity 
\mbox{$v_4\{\Psi_2\}/v_4\{\Psi_4\}=\mean{\cos{4(\Psi_2-\Psi_4)}}$}~\cite{Yan:2015jma}.
There are two trends common to all $n$ in \fig{fig:1}: (1)~in the 
low-\pt region the anisotropy appears largest for the lightest 
hadron and smallest for the heaviest hadron and (2)~in the 
intermediate-\pt~($3 \alt p_T\alt 4$~GeV/$c$) region this mass 
dependence partly reverses, such that the anisotropy is greater 
for the baryons ($N_q=3$) than for the mesons ($N_q=2$) at 
the same \pt. These trends remain significant after taking into 
account the \pt-correlated systematic uncertainties. These 
patterns have been observed previously in $v_2\{\Psi_2\}$ measurements for 
identified particles in Au$+$Au collisions at 
RHIC~\cite{Adler:2001nb,Adare:2012vq}, and are also seen here to 
hold for the higher moments $v_3\{\Psi_3\}$, $v_4\{\Psi_4\}$, and $v_4\{\Psi_2\}$. 
The mass dependence in the low-\pt range is a generic feature of 
hydrodynamical models, reflecting the mass ordering from the 
common velocity field ({\em i.e.} radial flow), and the 
dependence on valence quark number in the intermediate-\pt region 
has been associated with the development of flow in the partonic 
phase~\cite{Adare:2006ti}.

\begin{figure*}[tbp]
\includegraphics[width=0.998\linewidth]{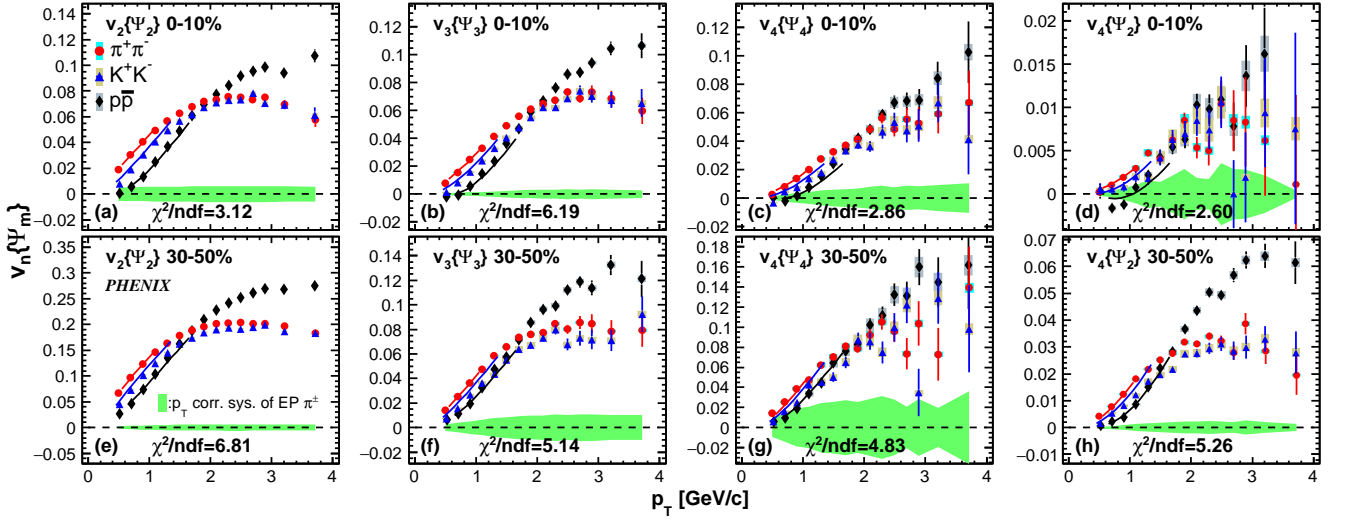}
\caption{(Color online) 
Fourier coefficients for charge-combined $\pi^{\pm}$, $K^{\pm}$, and 
$p+\bar{p}$ at midrapidity in Au$+$Au collisions at $\sqrt{s_{_{NN}}}$ = 
200 GeV. Coefficients are determined using the event plane method.
The curves illustrate the fits from the BW model.
Systematic uncertainties are shown as in \fig{fig:1}.
}
\label{fig:2}
\end{figure*}


{\it Results for finer centrality bins}.
The $v_n\{\Psi_m\}$ of $\pi^{\pm}$, $K^{\pm}$, and $p+\bar{p}$ measured with the 
event plane method are shown in \fig{fig:2} for the centrality selections
0\%--10\% and 30\%--50\%.
The same mass dependence of $v_n\{\Psi_m\}$ is seen in the low-\pt region for all harmonics and 
centralities. The evolution of baryon-meson splitting at intermediate-\pt 
is also observed for all centralities in $v_2\{\Psi_2\}$ and $v_3\{\Psi_3\}$ but could not be 
confirmed for $v_4\{\Psi_4\}$ in the most central and more peripheral events, or for 
$v_{4}\{\Psi_{2}\}$ in the most central events owing to the lower 
statistical significance of the measurements in those bins.

\begin{figure}[htbp]
\includegraphics[width=1.0\linewidth]{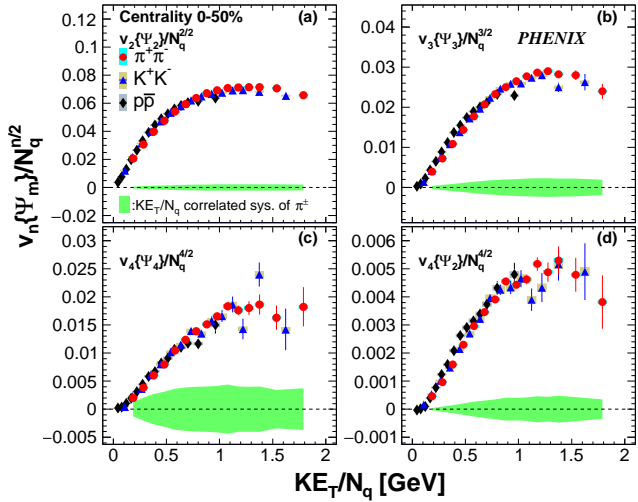}
\caption{(Color online) 
Quark-number ($N_q$) scaling for 0\%--50\% central Au$+$Au collisions at 
$\sqrt{s_{_{NN}}}$ = 200 GeV, 
where  $N_q$ is the constituent valence quark number of each hadron. 
Systematic uncertainties are shown as in Fig.~\ref{fig:1}.
}
\label{fig:3}
\end{figure}


{\it Quark-number scaling}.
The baryon-meson splitting in the intermediate-\pt region can be 
taken as an indication that the number of constituent valence 
quarks $N_q$ is an important determinant of final-state hadron 
flow in this range. Indeed, the $v_2\{\Psi_2\}$ data for identified hadrons 
had previously been seen to scale such that $v_{2}\{\Psi_2\}/N_q$ was the 
same for different particle species when evaluated at the same 
transverse kinetic energy per constituent quark number in the 
range $\KEt/N_q \alt 1$~GeV (\mbox{$\KEt\equiv m_T-m_0$} and 
\mbox{$m_T\equiv\sqrt{\pt^2+m_0^2}$}, where $m_0$ is the hadron 
mass) {\em i.e.} ``quark-number scaling"~\cite{Adare:2006ti,Adare:2012vq}.  
We have found that the present data obey a generalization of 
this scaling~\cite{Han:2011iy}, where for each harmonic order 
$n$, the values of $v_n\{\Psi_m\}/(N_q)^{n/2}$ vs $\KEt/N_q$ lie on a 
single curve for all the measured species within a $\pm15$\% 
range.  Figure~\ref{fig:3} shows the adherence of the data to 
this empirical scaling, which reflects the combination of quark-number 
scaling for $v_2\{\Psi_2\}$ by quark coalescence~\cite{Kolb:2004gi} 
and the empirical observation 
$v_n\{\Psi_n\}(p_T){\propto}(v_2\{\Psi_2\}(p_T))^{n/2}$~\cite{ATLAS:2012at}.  
Any explanation of the 
underlying physics needs to match this scaling over this \KEt 
range, and neither 
hydrodynamics~\cite{Ollitrault:1992bk,Luzum:2010ae,Borghini:2005kd,Gardim:2012yp}, 
nor naive quark coalescence alone~\cite{Zhang:2015skc} predicts 
this scaling for the higher moments.
It is notable that for $v_2\{\Psi_2\}$, there are 
deviations from valence-quark scaling at higher $p_T$ with mesons 
and baryons having comparable anisotropies~\cite{Adare:2012vq}.  
Reconciling the different physics as a function of $p_T$ remains 
an outstanding challenge.


{\it Blast-wave fitting}.  The BW 
model~\cite{Schnedermann:1993ws,Huovinen:2001cy,Adler:2001nb,Masui:2007sca} 
is a description of a fluid freeze-out state characterized by its 
temperature $T_{f}$ and its $\phi$-averaged maximal radial flow 
rapidity $\rho_{0}$.   Here we extend the BW description to incorporate 
azimuthal anisotropies in both radial rapidities $\rho_{n}\{\Psi_m\}$ and spatial 
density $s_{n}\{\Psi_m\}$ for $n=2,3,4$, using the empirically 
defined quantities \mbox{$\rho(n, m, \phi ,r) 
=\rho_0(1+2\rho_n\{\Psi_m\}\cos{(n\phi)})\times r/R^{\rm max}$} 
and \mbox{$S(n,m,\phi)=1+2s_{n}\{\Psi_m\}\cos{(n\phi)}$}. The 
spectra and anisotropies of all hadrons freezing out of the fluid 
can then be predicted via~\cite{Huovinen:2001cy,Adler:2001nb}

\begin{eqnarray}
\frac{dN}{p_{T}dp_{T}} &\propto& \int_{\ }^{R^{\rm max}} rdr \int d\phi \,
m_{T} I_0(\alpha_{t}) K_1(\beta_{t}) \label{eq:BW}, \\
v_{n}\{\Psi_m\}
&=&\frac{\int^{R^{\rm max}} rdr\int d\phi\cos{(n\phi)}
I_n(\alpha_{t})K_1(\beta_{t})S(n,m,\phi)}{\int_{\ }^{R^{\rm max}}
rdr  \int d\phi \, I_0(\alpha_{t}) K_1(\beta_{t})
S(n,m,\phi)},  \quad
  \nonumber
\end{eqnarray}

\noindent
where $I_n$ and $K_1$ are modified Bessel functions of the 
first and second kind, \mbox{$\alpha_t$ = $(\pt/T_f)$ $ 
\sinh{\rho(n, m, \phi ,r)}$}, and \mbox{$\beta_t$ = 
$(m_T/T_f)$ $\cosh{\rho(n, m, \phi ,r)}$}. Using single 
particle spectra from~\cite{Adler:2003cb} together with the 
present $v_n\{\Psi_m\}$ data, BW parameters $T_{f}$, 
$\rho_{0}$, $\rho_{n}\{\Psi_m\}$, and $s_{n}\{\Psi_m\}$ are 
extracted via simultaneous fitting of the $\pi^{\pm}$, 
$K^{\pm}$, and $p+\bar{p}$ data with a 
minimization of global $\chi^2$, separately for each 
centrality selection and each $v_n\{\Psi_m\}$.
The fit ranges used for the $\pi^{\pm}$, $K^{\pm}$, and $p+\bar{p}$ are $0.5<\pt<1.1$ GeV/$c$, $0.4<\pt<1.3$ GeV/$c$, and $0.6<\pt<1.7$ GeV/$c$, respectively.
The BW fits to $v_n\{\Psi_m\}(\pt)$+spectra are compared to the data in Fig.~\ref{fig:2}
for 0\%--10\% and 30\%--50\% central collisions, together with the global $\chi^2/ndf$ of the fits
determined using the quadrature sum of the statistical and systematic uncertainties of the data.
The global $\chi^2/ndf$ in 10\%--20\% and 20\%--30\% central collisions is similar to that in 0\%--10\% and 30\%--50\% central collisions.

\begin{figure}[htbp]
\includegraphics[width=1.0\linewidth]{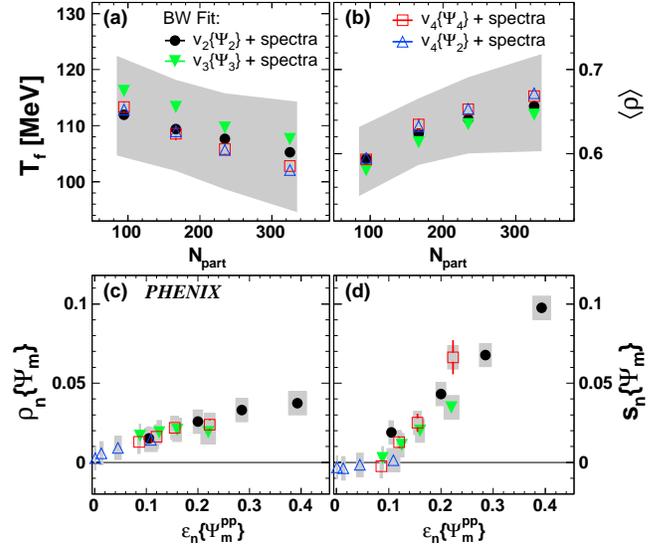}
\caption{(Color online) 
BW model fit parameters 
extracted for each $v_n\{\Psi_m\}$+spectra across different centrality classes.
The gray bands in (a)--(b) and shaded 
boxes in (c)--(d) indicate systematic uncertainties on the fitting \pt 
range and those propagated from the measurements. The width of the shaded 
boxes in $\varepsilon_{n}\{\Psi_m^{\rm PP}\}$ direction in
(c)--(d) indicates systematic uncertainties from Glauber models. Systematic uncertainties 
in (a) and (b) are similar among different fittings.
}
\label{fig:4}
\end{figure}

The results for the BW parameters are shown in Fig.~\ref{fig:4}. 
The freeze-out temperatures $T_f$ and radially averaged flow rapidities 
\mbox{$\left<\rho\right>=\int \left[\rho_0\times r/R_{\rm max}\right]rdr/\int rdr$} are in good agreement for the fits at different 
$n$, as would be required for a model of freeze-out.
$T_f$ and $\left<\rho\right>$ are primarily determined by the single particle spectra~\cite{Adler:2004hv},
while $\rho_n\{\Psi_m\}$ and $s_n\{\Psi_m\}$ are determined by $v_n\{\Psi_m\}$ measurements including \pt and particle mass dependences. 

The radial rapidity and spatial density anisotropies $\rho_n\{\Psi_m\}$ and $s_n\{\Psi_m\}$ 
extracted from the fits are shown against the average initial-state 
spatial participant-plane (PP) anisotropy 
\mbox{$\varepsilon_{n}\{\Psi_m^{\rm PP}\}=
\mean{\{r^2\cos{n(\phi^{\rm part}-\Psi_m^{\rm PP})}\}/\{r^2\}}$}, 
where $r$ and $\phi^{\rm part}$ are the 
polar coordinate positions of collision participant nucleons defined by 
Glauber models~\cite{Miller:2007ri,Alver:2010gr}, and $\Psi_m^{\rm PP}$ 
is the angle determined as 
\mbox{$\tan{(m\Psi_m^{\rm PP})}=\{r^2\sin{m\phi^{\rm part}}\}/\{r^2\cos{m\phi^{\rm part}}\}$}. 
Here, the brackets $\left<\right>$ and $\left\{\right\}$ 
denote averages over events and participants, respectively.  
The amplitude of $\varepsilon_n\{\Psi_m^{\rm PP}\}$ is smallest for the
most-central collisions and increases with centrality percentile. 

{\it Eccentricity of the medium at freeze out}. The $\rho_{n}\{\Psi_m\}$ 
and $s_n\{\Psi_m\}$ are generally smaller than the 
$\varepsilon_n\{\Psi_m^{\rm PP}\}$. The $\rho_{n}\{\Psi_m\}$ has a 
positive finite value and generally follows a common increasing curve as 
a function of $\varepsilon_n\{\Psi_m^{\rm PP}\}$ for $n=2,3,4$.  The 
$s_2\{\Psi_2\}$, $s_3\{\Psi_3\}$, and $s_4\{\Psi_4\}$ also show a common 
increasing trend in $\varepsilon_n\{\Psi_m^{\rm PP}\}\agt0.1$. We can 
interpret relative oscillations of event-plane dependent 
Hanbury-Brown-Twiss (HBT) radii with respect to averaged radii as the 
eccentricity of the medium at freeze-out if the direction of the radii 
is selected perpendicular to beam and pair momentum ($R_{\rm side}$), 
where these radii are less influenced by the emission duration and 
position-momentum correlations~\cite{Adare:2014vax}.


{\it Spatial information}.
Finite final eccentricities for $n=2$ and $n=3$ are observed 
by both the BW fit to $v_n\{\Psi_m\}$ and the event plane 
dependent HBT radii measurements using positive and negative 
pion pairs~\cite{Adare:2014vax}.
The $s_n\{\Psi_m\}$ therefore could reflect physical effects at the freeze-out of the medium.
The finite $s_n\{\Psi_m\}$ could be interpreted as a residual effect of 
initial state anisotropy $\varepsilon_n\{\Psi_m^{\rm PP}\}$, especially the 
contribution of initial-state fluctuations for $n=3,4$, after its 
dilution by the medium expansion. 
For $\varepsilon_n\{\Psi_m^{\rm PP}\}\alt0.1$, 
$s_3\{\Psi_3\}$, $s_4\{\Psi_4\}$, and $s_4\{\Psi_2\}$ are consistent with 
zero within systematic uncertainties. Comparisons of these small 
$s_n\{\Psi_m\}$ to the finite $\rho_n\{\Psi_m\}$ and $v_n\{\Psi_m\}$ in 
this $\varepsilon_n\{\Psi_m^{\rm PP}\}$ range
indicate that the anisotropic expansion velocity $\rho_n\{\Psi_m\}$ 
is a dominant source of the observed $v_n\{\Psi_m\}$ for higher harmonics. 
We expect this spatial information could provide new insights into
freeze-out conditions in hydrodynamic calculations.


{\it Summary and conclusions}.
In summary, the anisotropy strengths $v_2\{\Psi_2\}$, $v_3\{\Psi_3\}$, $v_4\{\Psi_4\}$, and 
$v_4\{\Psi_2\}$ for $\pi^{\pm}$, $K^{\pm}$, and $p+\bar{p}$ produced at 
midrapidity in Au$+$Au collisions at RHIC have been presented. The 
higher-order harmonics $v_n\{\Psi_m\}$ show particle mass splitting at low-\pt and 
baryon-meson difference at intermediate-\pt, very similar to what has been 
seen already for $v_2\{\Psi_2\}$.  The anisotropies obey a modified quark number 
scaling, where $v_n\{\Psi_m\}/(N_q)^{n/2}$ falls on a common trend against 
$\KEt/N_q$ for each $n$.  The data can be fit with a generalized BW model with 
empirically defined anisotropies in radial rapidity and spatial density at 
higher harmonic orders, which could provide a geometrical view of the 
hydrodynamical expansion at the end of freeze out.
Future analyses combining the results in this letter with
similar results from HBT and jet-like correlations with
respect to higher-order event planes
will further constrain the conditions and properties of the matter created 
at RHIC.




{\it Acknowledgments}.
We thank the staff of the Collider-Accelerator and Physics
Departments at Brookhaven National Laboratory and the staff of
the other PHENIX participating institutions for their vital
contributions.  We acknowledge support from the 
Office of Nuclear Physics in the
Office of Science of the Department of Energy, 
the National Science Foundation, 
Abilene Christian University Research Council, 
Research Foundation of SUNY, and 
Dean of the College of Arts and Sciences, Vanderbilt University (U.S.A),
Ministry of Education, Culture, Sports, Science, and Technology
and the Japan Society for the Promotion of Science (Japan),
Conselho Nacional de Desenvolvimento Cient\'{\i}fico e
Tecnol{\'o}gico and Funda\c c{\~a}o de Amparo {\`a} Pesquisa do
Estado de S{\~a}o Paulo (Brazil),
Natural Science Foundation of China (P.~R.~China),
Ministry of Education, Youth and Sports (Czech Republic),
Centre National de la Recherche Scientifique, Commissariat
{\`a} l'{\'E}nergie Atomique, and Institut National de Physique
Nucl{\'e}aire et de Physique des Particules (France),
Bundesministerium f\"ur Bildung und Forschung, Deutscher
Akademischer Austausch Dienst, and Alexander von Humboldt Stiftung (Germany),
National Science Fund, OTKA, K\'aroly R\'obert University College,
and the Ch. Simonyi Fund (Hungary),
Department of Atomic Energy and Department of Science and Technology (India), 
Israel Science Foundation (Israel), 
Basic Science Research Program through NRF of the Ministry of Education (Korea),
Physics Department, Lahore University of Management Sciences (Pakistan),
Ministry of Education and Science, Russian Academy of Sciences,
Federal Agency of Atomic Energy (Russia),
VR and Wallenberg Foundation (Sweden), 
the U.S. Civilian Research and Development Foundation for the
Independent States of the Former Soviet Union, 
the US-Hungarian Fulbright Foundation for Educational Exchange,
and the US-Israel Binational Science Foundation.


%
 
\end{document}